\documentclass{aa}
\usepackage{graphics}
\def\dst {\displaystyle}
\def\cm {{\rm cm}}
\def\gmcs{{\rm gm}~{\rm cm}^{-3}}
\def\K   {{\rm K}}

\def\kms {{\rm km}~{\rm s}^{-1}}
\def\hz  {{\rm Hz}}
\def\AA  {{\cal A}}
\def\abf {\alpha_{\rm bf}(\nu,i)}
\def\aT  {a_{\rm T}}
\def\Ba  {{\cal B}_a}
\def\Be  {{\cal B}_{\rm e}}
\def\CA  {{\bf C}_0}
\def\CB  {{\bf C}_1}
\def\CC  {{\bf C}_2}
\def\Ca  {{\cal C}_a}
\def\Ce  {{\cal C}_{\rm e}}
\def\cH  {\chi_{\rm H}}
\def\Cik {C_{i{\rm k}}}
\def\Cki {C_{{\rm k}i}}
\def\Dmb {\Delta m_{j-1/2}}
\def\Dmc {\Delta m_{j+1/2}}
\def\df  {\mbox{${\nabla\!\!\cdot\!{\bf F}_{\rm r}}$}}
\def\drf {\frac{1}{\rho}{\mbox{$\nabla\!\!\cdot\!{\bf F}_{\rm r}$}}}
\def\dVdt{{dV\over dt}}
\def\EI  {E_I}
\def\Ea  {E_a}
\def\Ee  {E_{\rm e}}
\def\Eex {E_{\rm ex}}
\def\me  {m_{\rm e}}
\def\mH  {m_{\rm H}}
\def\ne  {n_{\rm e}}
\def\nH  {n_{\rm H}}
\def\ni  {n_i}
\def\Pa  {P_a}
\def\Pe  {P_{\rm e}}
\def\Pg  {P_{\rm g}}
\def\Pik {P_{i{\rm k}}}
\def\Pki {P_{{\rm k}i}}
\def\Qea {Q_{{\rm e}a}}
\def\Qei {Q_{\rm ei}}
\def\Qelc{Q_{\rm elc}}
\def\Qinc{Q_{\rm inc}}
\def\Rik {R_{i{\rm k}}}
\def\Rki {R_{{\rm k}i}}
\def\Ta  {T_a}
\def\Te  {T_{\rm e}}

\def\trad{t_{\rm phi}}
\def\xH  {x_{\rm H}}
\def\Y   {{\bf Y}}
\def\Er  {E_{\rm rad}}
\def\FLy {F_{\rm Ly}}
\def\Fr  {F_{\rm r}}
\def\Dtj {\Delta\tau_{\nu j}}
\def\NC  {N_{\rm C}}
\def\Nm  {N_{\rm m}}
\def\NQ  {N_{\rm Q}}
\def\nuL {\nu_{01}'}
\def\Pr  {P_{\rm rad}}
\def\u   {u_{\mu\nu}}
\def\v   {v_{\mu\nu}}

\def\FrJb{\Fr^-}
\def\FrJc{\Fr^+}
\def\icr {\omega}
\def\neJb {\ne^-}

\def\nHJb {\nH^-}

\def\PgJb {\Pg^-}

\def\TaJb {\Ta^-}
\def\TaJc {\Ta^+}
\def\TeJb {\Te^-}
\def\TeJc {\Te^+}
\def\UJb  {U^-}

\def\XJ  {X_{\rm J}}
\def\XJb {X_{{\rm J}-1/2}}
\def\XJc {X_{{\rm J}+1/2}}

\begin{document}

\thesaurus{02         
           (02.19.1;  
            02.08.1;  
            02.18.7); 
         08(08.01.3)} 
\title{The Structure of Radiative Shock Waves}

\subtitle{I. The method of global iterations}

\author{Yu.A.~Fadeyev\inst{1}
        \and
        D.~Gillet\inst{2}}

\institute{Institute for Astronomy of the Russian Academy of Sciences,
 Pyatnitskaya 48, 109017 Moscow, Russia \\
 email: fadeyev@inasan.rssi.ru
 \and
 Observatoire de Haute-Provence - CNRS, F-04870
 Saint-Michel l'Observatoire, France \\
 email: gillet@obs-hp.fr }

\date{Received August 1997; accepted January 1998}

\titlerunning{The Structure of Radiative Shock Waves. I}
\maketitle

\begin{abstract}
The structure of steady plane-parallel radiative shock waves propagating
through the hydrogen gas undergoing partial ionization and excitation of
bound ato\-mic states is investigated in terms of the self-consistent
solution of the equations of fluid dynamics, radiation transfer and atomic
kinetics. The shock wave model is represented by a flat finite slab with no
incoming radiation from external sources at both its boundaries.
The self-consistent solution is obtained using the global iteration procedure
each step of which involves (1) integration of the fluid dynamics and
rate equations for the preshock and postshock regions, consecutively,
both solutions being fitted by the Rankine-Hugoniot relations at the
discontinuous jump; (2) solution of the radiation transfer equation for the
whole slab.
The global iteration procedure is shown to converge to the stable solution
which allows for the strong coupling of the gas flow and the radiation 
field produced by this flow.
Application of the method is demonstrated for the shock waves
with upstream velocities of $15~\kms\le U_1\le 60~\kms$
(i.e. with upstream Mach numbers $2.3\le M_1\le 9.3$)
and the hydrogen gas of unperturbed temperature $T=3000\K$ and
density $\rho = 10^{-10}~\gmcs$.

\keywords{Shock waves -- Hydrodynamics -- Radiative Transfer
         -- Stars: atmospheres}

\end{abstract}


\section{Introduction}

Radiative shock waves belong to the conspicuous phenomena demonstrating
the tight interplay between hydrodynamic motions and the radiation field.
The role of this interplay is strongest in the low gas density flows, so
the shocks are of tremendous importance in astrophysics. They are
observed in a wide variety of astrophysical phenomena: nova and
supernova explosions, bright filaments in old supernova remnants,
accretion flows in protostellar clouds. Shock waves are detected also in
atmospheres of radially pulsating variables such as Cepheids, RR~Lyr, W~Vir,
RV~Tau and Mira type stars. Periodic shocks propagating through pulsating
atmospheres lead to the distention of outer atmospheric layers and to
the mass loss.

Importance of radiative shock waves attracted attention of many authors
but nevertheless the shock properties are explained quite well still
qualitatively (see, for example, Zel'dovich \& Raizer 1966;
Skalafuris 1968; Mihalas \& Mihalas 1984;
Liberman \& Velikovich 1986).
The principal difficulty in obtaining the correct quantitative description
of the shock wave structure
is that the model has to allow for the strong coupling between
the gas flow and the radiation field, both them being characterized by
substantial departures from LTE.
Solution of this problem encounters serious difficulties, so that
in immensely numerous studies available at present in the literature
the authors used various assumptions and simplifications
(e.g. local thermodynamic equilibrium, treatment of the radiation transfer
in diffusion approximation, neglecting the opacity in the Balmer continuum
etc.). Many of these assumptions were found later inadequate or
leading to uncertain conclusions.
For instance, Kogure (1962), Sachdev (1968) and Hill (1972)
used the LTE approximation which does not hold as emphasized later
Narita (1973). Considering the hydrogen gas, Whitney \& Skalafuris~(1963)
relaxed this assumption but incorrectly assumed that the postshock region
is transparent for all hydrogen continua.
Finally, Narita (1973) took into account the opacity in the both
Lyman and Balmer continua.
The most elaborate numerical modelling of radiative shock waves based on
the self-consistent solution of the equations of fluid dynamics,
radiative transfer and atomic level populations was done by Klein et al.
(1976, 1978). However, the coarse zonning $(\sim 10^6~\cm)$ did not allow
to authors to consider the detailed structure of the shock front including
the radiative precursor and the thermalization zone where the electron
temperature gradually equalizes with temperature of heavy particles.
Nevertheless, this approach was found to be enough for consideration of shock
dynamics in atmospheres of A-type stars because the radiative precursor
is not so important due to the high temperature of the unperturbed gas.

Because astrophysical shocks in stellar atmospheres propagate
through the partially ionized hydrogen gas, a substantial fraction
of photons produced within the wake are absorbed in the radiative precursor.
As was shown by Gillet \& Lafon (1984, 1990) the structure of the
radiative precursor is complex and should be treated with same degree of
approximation as the postshock region.
In their studies Gillet \& Lafon (1984, 1990) treated the radiative
transfer as an initial value problem which was solved using the
shooting method. The principal obstacle in such an approach is that
the transfer equation possesses a singularity in the postshock region
(Gillet et al. 1989). Indeed, application of the eigenvalue methods
for solution of the nongrey transfer problem is affected by exponentially
growing errors (Mihalas 1978).

One of the first attempts to obtain the self-consistent solution for
the shock wave structure was undertaken by Nelson \& Goulard (1969)
and Nelson (1973). They considered the shock waves propagating through
the argon-like gas with upstream Mach numbers of $M_1=18$ and $M_1=24$.
The continuity, momentum and energy equations were written in the
integral representation whereas the radiation transfer was treated in
the simplified formulation.
The studies of radiative shock waves in helium and nitrogen
done by Clarke \& Ferrari (1965), Farnsworth \& Clarke (1971)
and Foley \& Clarke (1973) seem to be the best among known in the literature.
The authors emphasized the crucial role of the radiation transfer
treatment and employed the formal solution of the transfer equation.
The self-consistent shock wave models were obtained
in these studies with iteration procedure.
Unfortunately, there is a problem of exponential factors when the
formal solution is applied for optically thick layers.

In this paper we present a new approach based on the iterative solution
of the equations of fluid dynamics, the rate equations and the radiation
transfer equation. The momentum equation, the energy and rate equations
are written in the form of ordinary differential equations.
These equations are stiff and such a representation is most appropriate
from the point of view of stability and small truncation errors.
The radiation transfer is treated as a two-point boundary value problem.
This allows us to obtain the stable solution of the transfer equation for the
whole spectral range including both the opaque Lyman continuum and
the more transparent higher order continua.
The method of global iterations takes into account the coupling between the
gas flow and the radiation field, so that the structure of the
radiative shock wave is considered in terms of the self-consistent model.

In the framework of this first approach, only devoted to provide a new
technique for obtaining the self-consistent solution, we consider the structure
of steady, plane-parallel shock waves propagating through an infinite, 
isotropic, pure hydrogen plasma. The steady assumption is correct, for example,
to a good degree of approximation in most applications to stellar atmospheres.
Indeed, the time required for the gas flow to cross the shock wake with
typical thickness of 10$^{4}$-10$^{7}$~cm is much less than the characteristic 
time during of which the shock wave energy appreciably decreases.
In pulsating stars, the radiative lifetime of a shock is between a few
hours and a few months, which, consequently, is much larger than the
10$^{-2}$-10~s of the gas flow time to cross the shock wake.
Accuracy of the plane-parallel approximation follows from the
very small width of the shock wave in comparison with stellar radius.

After describing the shock wave model (Sect.~2) we derive the system
of ordinary differential equations (Sect.~3). The radiation transfer
equation is solved for the whole shock wave model
using the Feautrier technique (Sect.~4).
In Sect.~5 we show that the global iteration procedure comprising the
initial value problem for ordinary differential equations and the two-point
boundary value problem for radiation transfer converges to the
self-consistent solution. Results of calculations demonstrating the
applicability of the method are given in Sect.~6. Finally, in Sect.~7,
we give some concluding remarks and discuss the future aspects of the problem.


\section{The shock wave model}

Consider a steady, plane-parallel shock wave propagating through the
homogeneous medium which is at rest and consists of a pure hydrogen gas.
No radiation and gravitational forces from external sources are assumed
to be present. The problem to be solved is that to describe the spatial
structure of the shock wave in terms of the self-consistent solution of the
equations of fluid dynamics, the rate equations for hydrogen atomic level
populations and the radiation transfer equation.
The problem is characterized by three input parameters:
the temperature $T_1$ and the density $\rho_1$ of the unperturbed gas
as well as the speed $U_1$ at which the gas material flows into the shock.

The structure of radiative shock waves is schematically divided into four
zones:
(1) a precursor, where the gas is heated and is partially ionized by
radiation emerging from the postshock region;
(2) a very narrow zone, where a major part of the kinetic energy of the
upstream flow is converted due to viscosity and conductivity into the thermal
energy of translational motions of heavy particles, that is, neutral atoms
and ions;
(3) a thermalization zone, where the kinetic energy of translational
motions of heavy particles is redistributed among various degrees of
freedom;
(4) a radiative relaxation zone, where hydrogen atoms recombine and the gas
radiatively cools.
Because of its extremely small width (a few mean free paths of gas particles),
the second zone cannot be correctly described in terms of the fluid dynamics
and, hence, should be considered as a discontinuous jump across which
the Rankine-Hugoniot equations are applied.
Thus, the present study is confined by consideration of the spatial structure
of the radiative precursor as well as the thermalization and relaxation zones.

Let the origin of the comoving frame to coincide with infinitesimaly thin
discontinuous jump dividing the me\-dium into the preshock and the
postshock regions. The spatial coordinate is $X=0$ at the discontinuous
jump, is negative in the preshock region and is positive in the
postshock region. The shock wave model is represented by a flat finite
slab comoving with discontinuous jump.
Thus, the velocity of the gas material flowing through the slab is
always positive: $U = dX/dt > 0$.
At the outer boundary of the
preshock region with spatial coordinate $X_1 < 0$ the gas is assumed
to be unperturbed. The spatial coordinate of the postshock outer boundary
is $X_N > 0$ and physical properties of the gas at this point are not known.

The radiation transfer equation is solved for the whole slab in the framework
of the two-point boundary value problem, therefore the slab is
represented by a set of spatial cells. The discontinuous jump locates
at the $J$-th cell boundary with spatial coordinate $\XJ=0$.
The cells are inequally spaced, are smallest at the discontinuous jump and
increase outwardly in both directions from the discontinuous jump
according to the geometrical progression.
Following the traditional conventions of computational radiation hydrodynamics
(see, for example, Mihalas \& Mihalas 1984),
all thermodynamic variables are defined at the cell centers
$X_{j-1/2} = \frac{1}{2}\left(X_{j-1}+X_j\right)$ and are denoted by
half-integer subscripts.
The spatial coordinates of the cell centers nearest to the discontinuous
jump hereafter are referred as the inner boundaries of the preshock and
postshock regions, respectively, and are denoted as $\XJb$ and $\XJc$.

The assumption of the steady shock wave allows us to reduce the equations
of fluid dynamics to a system of ordinary differential
equations. In order to take into account the coupling between hydrodynamic
motions and the radiation field we employ the iteration procedure comprising
the consecutive solution of the transfer equation and integration of the
rate and fluid dynamics equations.
The starting point for integration of ordinary differential equations is
the outer boundary of the preshock region $X_1$ where the gas is
assumed to be unperturbed. The preshock integration is done
within the interval $[X_1,\XJb]$.
Then we solve the Rankine-Hugoniot equations
\begin{eqnarray}
\label{C0}
&&\varrho U \equiv \dot m = \CA
\,,\\
\label{C1}
&&\dot m U + \Pg = \CB
\,,\\
\label{C2}
&&\dot m
\left({\dst{h\over\varrho}} + \dst{\frac{1}{2}}U^2\right) + \Fr = \CC
\,,
\end{eqnarray}
where $h$ is the enthalpy, 
\begin{equation}
\label{enth}
h=5/3E_{trans}+E_{rot}+E_{vib}+E_{el}+E_{diss}+E_{ion}
\end{equation}
where $E_{trans}$, $E_{rot}$, $E_{vib}$, $E_{el}$, $E_{diss}$ and $E_{ion}$
are the translational energy; rotational, vibrational, electronic excitation 
internal energies and dissociation and ionization potential energies 
respectively. In this paper only $E_{trans}$ and $E_{ion}$ are different of 
zero. $\Fr$ is the total radiative flux,
$\CA$, $\CB$ and $\CC$ are the mass, momentum and energy fluxes across
the discontinuous jump.
In the present study we assume that the radiation pressure $\Pr$ and
the radiation energy density $\Er$ can be neglected in comparison with
gas pressure and internal energy
of the gas material. Applicability of this assumption is shown below.

Let us designate, for the sake of convenience,
the quantities defined at $\XJb$ by the superscript minus 
and the quantities defined at $\XJc$ by the superscript plus.
Eqs.~(\ref{C0}) -- (\ref{C2}) are solved for the temperature of heavy
particles $\TaJc$ and the inverse compression ratio
$\omega=\rho^-/\rho^+$ $=U^+/U^-$.
Assuming that hydrogen atoms undergo across the discontinuous jump
neither excitation of the bound levels nor ionization and that
the hydrogen ions have the same temperature as that of neutral hydrogen
atoms, we obtain
%
\begin{eqnarray}
\label{TaJc}
\TaJc &=& \TaJb +
{\neJb\over\nHJb}\left(\TeJb-\TeJc\right) +
\frac{1}{5}{\dot m\UJb\over\nHJb k} \left(1-\omega^2\right) +
\nonumber\\
&+& \frac{2}{5}{\FrJb-\FrJc\over\UJb\nHJb k}
\,,
\end{eqnarray}
where $k$ is the Boltzmann constant, $\nH$ and $\ne$ are the total
numbers of hydrogen atoms and free electrons per unit volume,
respectively.
It should be noted that jump conditions (1) -- (3) are not applied to the
infinitasimally thin discontinuous jump coinciding with the cell interface
$\XJ$ but rather relate the physical variables at cell centers $\XJb$ and
$\XJc$. That is why the term containing ${\Fr}^- - {\Fr}^+$ into Eq.~(5)
does not cancel and has to be taken into account.
The inverse compression ratio is obtained from
\begin{equation}
\label{icr}
A\icr^2 + B\icr + C = 0
\,,
\end{equation}
where
\begin{eqnarray}
A &=& \frac{4}{5}\dot m\UJb
\,,\\
B &=& -\dot m\UJb - \PgJb
\,,\\
C &=& \frac{1}{5}\dot m\UJb + \PgJb + \frac{2}{5}{\FrJb-\FrJc\over \UJb}
\,.
\end{eqnarray}

Eqs.~(\ref{TaJc}) and (\ref{icr}) imply that the postshock electron
temperature $\TeJc$ at the cell center $\XJc$ is known.
Across the discontinuous jump the elctron gas undergoes the adiabatic
compression and the electron temperature increases by a factor of
$1/\omega^{\gamma -1}$, where $\gamma$ is the ratio of specific heats
(Zel'dovich \& Raizer 1966). Because the adiabatic compression has a weak
effect and the electron heat conduction is out the
scope of the present study, we assumed that the electron temperature
does not change across the discontinuous jump, that is, $\TeJb = \TeJc$.

Eqs.~(\ref{TaJc}) and (\ref{icr}) are solved to determine the initial
conditions for the postshock integration which is done within the interval
$[\XJc,X_N]$.
Integration of ordinary
differential equations provides with spatial distibutions of
electron temperature $\Te$, gas density $\rho$, the number
densities of free electrons $\ne$ and atomic level populations $\ni$.
These quantities are used in solution of the radiation
transfer equation which gives improved radiation intensities
within the whole slab.
The global iteration procedure consisting of the integration of ordinary
differential equations and solution of the radiation transfer equation
continues until the relative
changes of all quantities become less than the convergence criterion.
Below we discuss details of the global iteration procedure.


\section{The system of ordinary differential equations}

The solution vector to be found from integration of the system of
ordinary differential equations consists of $L+4$ variables and is
\begin{equation}
\Y = \left\{U,\ \Ea',\ \Ee',\ \ne',\ n_1',\ \ldots ,\ n_L' \right\}
\,,
\end{equation}
where
$L$ is the number of bound atomic states treated in non-LTE,
$U$ is the gas material velocity with respect to the discontinuous jump,
\begin{equation}
\Ea' = {\Ea\over\rho} = \frac{3}{2}{\nH k\Ta\over\rho}
\end{equation}
is the specific translational kinetic energy of heavy particles
(neutral atoms and ions),
\begin{equation}
\Ee' = {\Ee\over\rho} = \frac{3}{2}{\ne k\Te\over\rho}
\end{equation}
is the specific translational kinetic energy of free electrons,
$\ne'$ is the number of free electrons per unit mass,
$\ni'$ is the number of hydrogen atoms in the $i$-th state per unit mass.
Hereafter the prime implies that the quantity is expressed per unit mass.
The solution vector $\Y$ does not contain the gas density $\rho$
because this variable can be easily evaluated from the mass
conservation relation~(\ref{C0}).

Thus, the system of ordinary differential equations consists of the momentum
equation, two energy equations for heavy particles and free electrons,
the rate equation for free electrons and $L$ rate equations for
non-LTE bound states of the hydrogen atom:
\begin{eqnarray}
\label{dUdt}
{dU\over dt} &=& -\frac{1}{\rho}{d\Pg\over dX}
\,,\\
\label{dEadt}
{d\Ea'\over dt} &=& -\Pa\dVdt - \Qelc
\,,\\
\label{dEedt}
{d\Ee'\over dt} &=& -\Pe\dVdt + \Qelc + \Qinc
\,,\\
\label{dnedt}
{d\ne'\over dt} &=& \frac{1}{\rho}\sum\limits_{i=1}^L
\left(\ni\Pik - \ne\Pki\right)
\,,\\
\label{dnidt}
{d\ni'\over dt} &=& \frac{1}{\rho}
\left(\ne\Pki - \ni\Pik\right)
\,,
\qquad\left(i=1,\ldots ,L\right)
\,,
\end{eqnarray}
where
$\Pa=\nH k\Ta$ is the partial pressure of heavy particles,
$\Pe=\ne k\Te$ is the electron pressure,
$\Qelc$ and $\Qinc$ are the rates of energy gain by electrons in elastic
and inelastic collisions, respectively,
$\Pik$ and $\Pki$ are the total, that is collisional plus radiative,
ionization and recombination rates, respectively.
It should be noted that in the present study we consider only
bound-free transitions, so that equations (\ref{dnedt}) and (\ref{dnidt})
contain only ionization and recombination terms.

The system of ordinary differential equations (\ref{dUdt}) -- (\ref{dnidt})
written in the form of derivatives with respect to time $t$
is not appropriate for calculation of the spatial structure,
so that these equations should be rewritten in order their left-hand
sides are replaced by derivatives with respect to the spatial coordinate $X$.
Furthermore, the space derivative of the gas pressure $\Pg$ in
Eq.~(\ref{dUdt}) and the time derivative of the specific volume $V$
in Eqs.~(\ref{dEadt}) and (\ref{dEedt})
have to be expressed in terms of integrated variables.
To this end we write the gas pressure as a sum of translational kinetic
energies:
\begin{equation}
\label{Pg}
\Pg = \nH k\Ta + \ne k\Te = \frac{2}{3}\Ea + \frac{2}{3}\Ee
\,,
\end{equation}
whereas the time derivative of the specific volume is determined from
Eq.~(\ref{C1}):
\begin{equation}
\label{dVdt}
\dVdt = -{1\over\dot m^2}{d\Pg\over dt}
\,.
\end{equation}
Expressing the time derivative of the specific volume in terms 
of the gas pressure according to Eq.~(19), $\Pg$ according to
Eq.~(18) in terms of integrated quantities $\Ea'$ and $\Ee'$ and
substituting Eqs.~(18) and (19) into Eqs.~(13) -- (17), we obtain
the following system of ordinary differential equations
with right--hand sides depending only on the
independent variable $X$ and integrated variables
$\left\{U,\ \Ea',\ \Ee',\ \ne',\ n_1',\ \ldots ,\ n_L' \right\}$:
\begin{eqnarray}
\label{dUdX}
{dU\over dX} & = &
-{\AA\over U}\left(\Ba + \Be\right)
\,,\\
\label{dEadX}
{d\Ea'\over dX} & = &
\Ba\left(1-\Ce\right) + \Be\Ca
\,,\\
\label{dEedX}
{d\Ee'\over dX} & = &
\Ba\Ce + \Be\left(1 - \Ca\right)
\,,\\
\label{dnedX}
{d\ne'\over dX} & = & \sum\limits_{i=1}^L
{\ni\Pik - \ne\Pki\over\dot m}
\,,\\
\label{dnidX}
{d\ni'\over dX} & = &
{\ne\Pki - \ni\Pik\over\dot m}
\,,\qquad\left(i=1,\ldots ,L\right)
\,,
\end{eqnarray}
where
\begin{eqnarray}
\label{AA}
\AA &=& \frac{2}{3}{U^2\over U^2 - \aT^2}
\,,\\
\label{Ba}
\Ba &=& -{\Qelc\over U}{1-\beta^2\over 1-\frac{5}{3}\beta^2}
\,,\\
\label{Be}
\Be &=&
\left\{
{\Qelc\over U} -
\cH
\left[
{d\ne'\over dX} +
\sum\limits_{i=1}^L\left(1-\frac{1}{i^2}\right){d\ni'\over dX}
\right] -
\right.
\nonumber\\
&-&
\left.
{\df\over\dot m}
\right\}
{1-\beta^2\over 1-\frac{5}{3}\beta^2}
\,,\\
\label{Ca}
\Ca &=& {\AA\Pa\over U\dot m}
\,,\\
\label{Ce}
\Ce &=& {\AA\Pe\over U\dot m}
\,,
\end{eqnarray}
$\beta=\aT/U$,
$\aT = \sqrt{\Pg/\rho}$ is isothermal sound speed,
$\cH=13.598$~eV is the ionization potential of the hydrogen atom.

In obtaining Eq.~(\ref{Be}) we expressed the rate of energy gain by electrons
in inelastic collisions as (Murty 1971)
\begin{equation}
\label{Qinc}
\Qinc =  - {d\over dt}\left(\EI'+\Eex'\right) - \drf
\,,
\end{equation}
where $\EI'$ and $\Eex'$ are ionization and excitation energies
per unit mass,
\begin{equation}
\label{drf}
\df = 4\pi\int\limits_0^\infty\left(\eta_\nu - \kappa_\nu J_\nu\right)\,d\nu
\end{equation}
is the divergence of radiative flux,
$J_\nu$ is the mean intensity of radiation,
$\eta_\nu$ and $\kappa_\nu$ are the total emission and absorption coefficients.

Free electrons acquire the energy from heavy particles in elastic
collisions with hydrogen ions and neutral hydrogen atoms, hence,
\begin{equation}
\label{Qelc}
\Qelc = \Qei + \Qea
\,,
\end{equation}
where $\Qei$ and $\Qea$ are the corresponding rates of energy gain.
The rate of energy gain by electrons in elastic collisions with hydrogen
ions per unit mass is (Spitzer \& H\"arm 1953)
\begin{equation}
\label{Qei}
\Qei = \frac{2}{3}\frac{\ne}{\varrho}k{\Ta-\Te\over t_{eq}}
\,,
\end{equation}
where $t_{\rm eq}$ is the time of equipartition given by
\begin{equation}
\label{teq}
t_{\rm eq} = {252\Te^{3/2}\over\ne\ln\Lambda}
\end{equation}
and
\begin{equation}
\ln\Lambda = 9.43 + 1.15\log\left(\Te^3/\ne\right)
\,.
\end{equation}
The rate of energy gain by electrons in elastic collisions with neutral
hydrogen atoms per unit mass is
\begin{equation}
\label{Qea}
\Qea = {\ne\me\over\nH\mH}{\Ta-\Te\over\Te}
n_1\langle\sigma_{ea}v^3\rangle
\,,
\end{equation}
where $\me$ and $\mH$ are the mass of electron and the mass of hydrogen
atom, respectively, and the elastic scattering cross section is (Narita 1973)
\begin{eqnarray}
\langle\sigma_{ea}v^3\rangle =
\int\limits_0^\infty
\sigma_{ea}v^3f\left(v\right)\,dv =
\nonumber\\
=
4\pi a_0^2\left({8\over\pi}\right)^{1/2}
\left({k\Te\over\me}\right)^{3/2}
\left[
4 + {24\over
\left(1+2\cdot 10^{-5}\Te\right)^3}
\right].
\end{eqnarray}
Here $a_0$ is the Bohr radius.

Rate equations (\ref{dnedX}) and (\ref{dnidX}) imply that
the number density of free electrons $\ne$ and atomic level populations $\ni$
change due to bound-free transitions, that is, due to ionizations and
recombinations.
The total ionization rate is
\begin{equation}
\label{Pik}
\Pik = \ne\Cik + \Rik
\,,
\end{equation}
where the rate of collisional ionizations is given by
\begin{equation}
\label{Cik}
\Cik = \pi a_0^2
\left({8k\over\me\pi}\right)^{1/2}
\Te^{1/2}
\exp\left(-{\chi_i\over k\Te}\right)
\Gamma_i\left(\Te\right)
\,,
\end{equation}
$\chi_i = \cH/i^2$ is energy of ionization from the $i$-th level,
$\Gamma_i\left(T\right)$ is a slowly varying function of $T$ evaluated with
approximation formulae by Mihalas (1967).

The rate of photoionizations is
\begin{equation}
\label{Rik}
\Rik = 4\pi\int\limits_{\nu_{0i}}^\infty
{\abf\over h\nu}J_\nu\, d\nu
\,,
\end{equation}
where $\abf$ is an absorption cross-section at
frequency $\nu$ in bound-free transition from the $i$-th state and
$\nu_{0i}$ is a threshold frequency for ionization from the $i$-th state.

The total recombination rate is
\begin{equation}
\label{Pki1}
\Pki = \ne\Cki + \Rki
\,,
\end{equation}
where the collisional recombination rate is given by
\begin{equation}
\label{Cki}
\Cki = {\ni^*\over\ne^*}\Cik
\,.
\end{equation}
The radiative recombination rate is
\begin{equation}
\label{Rki}
\Rki ={\ni^*\over\ne^*}\Rik^\dagger
\,,
\end{equation}
where
\begin{equation}
\label{Rikd}
\Rik^\dagger =
4\pi\int\limits_{\nu_{0i}}^\infty
{\abf\over h\nu}
\exp\left(-{h\nu\over k\Te}\right)
\left({2h\nu^3\over c^2} +J_\nu\right)\, d\nu
\,.
\end{equation}
Substituting Eqs.~(\ref{Cki}) and (\ref{Rki}) into Eq.~(\ref{Pki1})
we obtain that the total recombination rate is
\begin{equation}
\label{Pki2}
\Pki = {\ni^*\over\ne^*}
\left(\ne\Cik + \Rik^\dagger\right)
\,.
\end{equation}

The system of ordinary differential equations (\ref{dUdX}) -- (\ref{dnidX}) 
is stiff because it is characterized by very different time constants
due to the rate equations (\ref{dnedX}) and (\ref{dnidX}).
In order to obtain the stable and enough correct solution of
Eqs.~(\ref{dUdX}) -- (\ref{dnidX}) we used
the Livermore solver for ordinary differential
equations based on the GEAR package (Hindmarsh 1979).


\section{The radiation transfer equation}

For the shock wave models considered in the present study the extinction
coefficient is highest at the Lyman edge frequency $\nu_{01}$.
In the preshock region the corresponding extinction coefficient is
$\chi(\nu_{01})\sim 3\cdot 10^{-4}~\cm^{-1}$.
Behind the discontinuous jump the extinction coefficient initially
decreases because of ionization of hydrogen atoms and then increases
within the recombination zone.
For all the models considered
$\chi(\nu_{01})\la 5\cdot 10^{-3}~\cm^{-1}$.
Thus, the time needed for photons to travel a mean free path is
$t_{\rm ph} = [c\chi(\nuL)]^{-1} \sim 7\cdot 10^{-9}$~s,
where $c$ is the velocity of light.
Because a characteristic structural length for Lyman photons is only a few
mean free paths, a photon flight time
is much shorter than the characteristic time of producing the
changes due to hydrodynamic motions and, therefore,
the time-derivative term in the transfer equation can be omitted
because the radiation field is quasi static to a good accuracy.

In plane-parallel geometry the quasi static radiative transfer equation
is written as
\begin{equation}
\label{rte}
\mu{dI_\nu\over d\tau_\nu} = I_\nu - S_\nu
\,,
\end{equation}
where $I_\nu$ is the specific intensity of radiation depending on the
frequency $\nu$, the directional cosine $\mu$ and the mono\-chromatic
optical depth $d\tau_\nu = \chi_\nu dX$.

The present study is confined by treatment of the continuum radiation transfer
with total extinction coefficient given by
\begin{equation}
\label{chi}
\chi_\nu = \kappa_{\rm bf}(\nu) + \kappa_{\rm ff}(\nu) +
\sigma_{\rm T}\ne + \sigma_{\rm R}(\nu)n_1
\,,
\end{equation}
where $\kappa_{\rm bf}(\nu)$ and $\kappa_{\rm ff}(\nu)$ are bound-free
and free-free absorption coefficients, respectively.
They can be found, for instance, in Mihalas (1978).
$\sigma_{\rm T}=6.65\cdot 10^{-25}~\mbox{cm}^2$ is the Thomson scattering
cross section
and $\sigma_{\rm R}(\nu)$ is the Rayleigh scattering cross section
evaluated due to approximate formulae by Kurucz (1970).
Because the scattering is assumed to be isotropic, the source function
$S_\nu$ can be written as
\begin{equation}
\label{Sf}
S_\nu = {\eta_\nu\over\chi_\nu}
\,.
\end{equation}
The source function $S_\nu$ is determined from integration of the system of
ordinary differential equations (\ref{dUdX}) -- (\ref{dnidX}) and is
evaluated at each cell center $X_{j-1/2}$ as a function of frequency $\nu$.

Solution of the radiation transfer equation when applied to the radiative
shock waves is accompanied by some difficulties. First, because the shock
wave propagates in the nearly neutral hydrogen gas, the slab has the
appreciably large optical depth in the Lyman continuum $(\tau\sim 10^2)$,
whereas in the
Balmer and higher order continua the total optical depth is very small.
For example, the ratio of the total optical depth at the Lyman continuum edge
to that in the Balmer continuum is as high as $\sim 10^6$.
Thus, in order to obtain the stable solution for the entire frequency range
we have to treat the transfer equation as a two-point boundary value problem.
Second, the optical depth increments
\begin{equation}
\label{Dtj}
\Dtj = \frac{1}{2}
\left(\omega_{\nu j-1/2}\Dmb + \omega_{\nu j+1/2}\Dmc\right)
\,,
\end{equation}
are extremely small for hydrogen continua of order $i\ge 3$.
Here $\omega_\nu = \chi_\nu/\rho$ and
$\Dmb = \rho_{j-1/2}\Delta X_{j-1/2}$ is the column mass contained in the cell.
The straightforward application of the Feautrier method fails
because of the limited machine accuracy. To alleviate this obstacle we
employ an improved Feautrier solution proposed by Rybicki \& Hummer (1991)
and providing with much better numerical conditioning of the recurrence
elimination scheme.

Thus, the transfer equation (\ref{rte}) is transformed (see, for example,
Mihalas 1978) into the second-order differential equation
\begin{equation}
\label{rte2}
\mu^2{d^2\u\over d\tau_\nu} = \u - S_\nu
\,,
\end{equation}
where $\u$ is a mean-intensity-like variable defined by
\begin{equation}
\label{u}
\u = \frac{1}{2}\left[I(\mu,\nu) + I(-\mu,\nu)\right] 
\,,
\end{equation}
and $\mu$ changes in the range $0\le\mu\le 1$.

The transfer equation (\ref{rte2}) is subject to boundary conditions at
both surfaces of the slab. Assuming that radiation is produced only by
the shock wave and that there is no incoming radiation from external sources,
we have
%
\begin{equation}
\label{bc}
\begin{array}{lll}
& I_\nu(\mu) = 0   & \quad\mbox{for\ } X=X_1 \,,\\
& I_\nu(-\mu) = 0  & \quad\mbox{for\ } X=X_N \,,
\end{array}
\end{equation}
The boundary conditions (\ref{bc}) are used in the second-order
accuracy Taylor's expansion of the mean-intensity-like variable $u_\nu$
at both boundaries of the slab and are sufficient to complete the system
of the finite-difference transfer equations.

The transfer equation (\ref{rte2}) is solved each cycle of global iterations
for mean-intensity-like variable $\u$ defined at the cell centers.
The total number of cells is $500\le N\le 1200$ depending on the shock wave
model.
The frequency range $\nu_B\le\nu\le\nu_A$ is divided into $\NC$ intervals,
where the upper boundary of the range is $\nu_A = 10^{16}~\hz$.
Boundaries of the intervals correspond to the threshold
ionization frequencies and the lower boundary of the frequency range
$\nu_B$ is the ionization threshold frequency of the $\NC$-th bound state.
Within each interval the integral with respect to frequency $\nu$
is replaced by the Gaussian quadrature sum, so that the integral
over the whole frequency range $[\nu_B,\nu_A]$ is obtained by summation
of interval integrals.
The angular range $0\le\mu\le 1$ is also replaced by a set of angular points
$\{\mu_i\}$ at the Gaussian quadrature nodes.
In the present study the number of quadrature nodes within each frequency
interval and the number of angular points were $\NQ=\Nm=4$.
Integrating $u_\nu$ with respect to $\mu$ and $\nu$ we evaluated in each
cell center the mean intensity $J_\nu$, the radiation energy density $\Er$,
the radiation pressure $\Pr$
and the divergence of radiative flux $\df$.
The total radiative flux was calculated in cell boundaries from
\begin{equation}
\Fr = 4\pi\int\limits_0^\infty\int\limits_0^1\v\mu\:d\mu\:d\nu
\,,
\end{equation}
where
\begin{equation}
\v = \frac{1}{2}\left[I(\mu,\nu) - I(-\mu,\nu)\right]
   = \mu{d\u\over d\tau_\nu} 
\end{equation}
is an antisymmetric flux-like variable.
It should be noted that according to our conventions the radiative flux
is negative if the radiation propagates in negative direction.
Thus, the radiative flux emerging ahead the discontinuous jump is always
negative.


\section{The global iteration procedure}

\begin{figure}
\resizebox{\hsize}{!}{\includegraphics{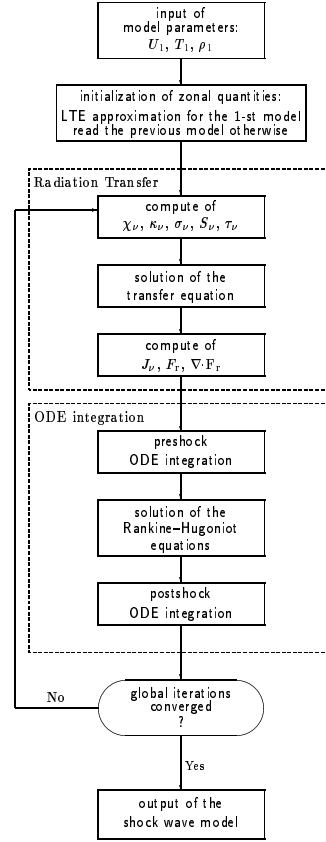}}
\caption{The flow chart of the global iteration procedure}
\label{fchrt}
\end{figure}

For integration of the system of ordinary differentail equations
(\ref{dUdX}) -- (\ref{dnidX}) we have to know the mean intensity $J_\nu$
and the divergence of radiation flux $\df$ at each cell
center of the shock wave model. On the other hand, 
the radiation transfer equation (\ref{rte2}) can be solved only when
the extinction and emission coefficients are given in each cell center.
In order to take into account a coupling between gas material
and radiation field we employ the iteration procedure.
If the initial approximation is enough close to the final solution,
we may hope that each iteration will give a better
approximation for the final solution than the previous one.
In Fig.~\ref{fchrt} is shown the flow chart where the main steps of
the global iteration procedure are depicted.

Within the whole shock wave both the radiation field and atomic level
populations are in a strong departure from LTE. This feature is the
principal difficulty accompanying the shock wave model calculations because
it is responsible for the narrow convergence area of global iterations.
In particular, the use of the initial LTE approximation
allows the converged solution to be obtained only for the weak shock waves
with upstream Mach numbers $M_1\la 3$. For larger upstream
velocities the initial oscillation amplitude of the solution vector $\Y$
becomes so large that some quantities fall beyond their physical meaning.
In order to alleviate this difficulty and
to be able to consider the structure of stronger shock waves
we computed a grid of the shock wave models with
gradually increasing upstream Mach number, the LTE initial approximation
being used only for the first model with the Mach number $M_1 = 2.3$.
Thus, each model of the grid with exception of the first one
was computed with initial approximation obtained from the previous converged model
having somewhat smaller upstream velocity. The upstream velocity increment
was in the range $1~\kms\le\Delta U_1\le 5~\kms$.
For test purposes some models were computed with different initial
conditions taken from shock wave models with both larger and smaller
upstream velocities. In all such cases the shock wave structure was found
to converge to s single solution.

\begin{figure}
\resizebox{\hsize}{!}{\includegraphics{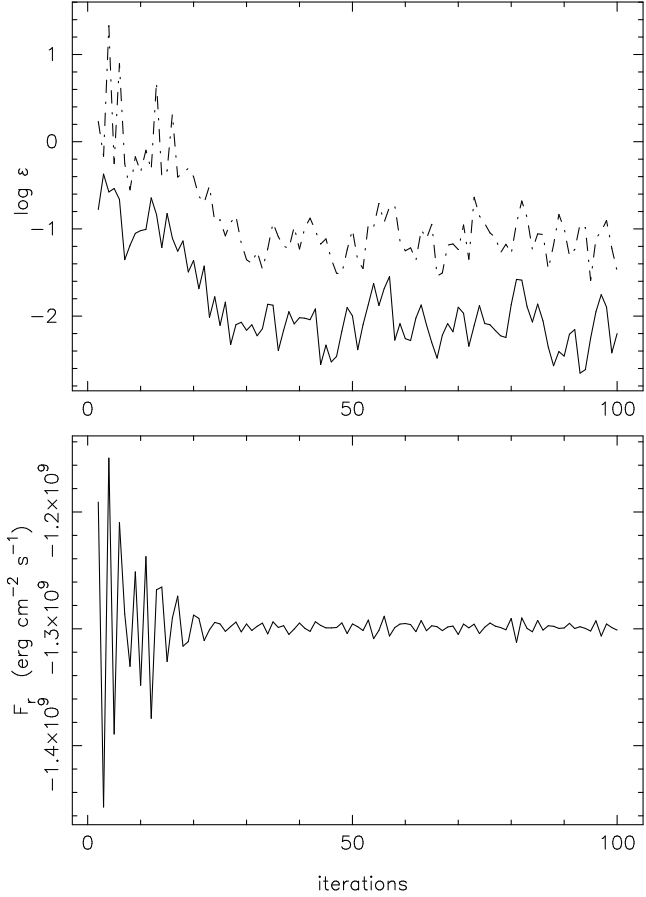}}
\caption{Upper panel: convergence plots for the electron temperature
(solid line) and the divergence of radiative flux (dot-dashed line)
given by the maximum absolute relative change $\varepsilon$.
Lower panel: convergence plot for the total radiative flux emerging from the
outer boundary of the preshock region}
\label{itrh}
\end{figure}
%

In order to describe the convergence of global iterations we introduce
for each component of the solution vector $\Y$ at the $\ell$-th
iteration the quantity
$$
\epsilon = \max_{1\le j\le N}
\left\vert 1 - {y_j^{\ell -1}\over y_j^\ell}\right\vert
$$
specifying the maximum absolute relative change of the variable $y$
across the whole shock wave model.
A typical behaviour of global iterations is shown
in Fig.~\ref{itrh} displaying in the upper panel
the maximum absolute relative change of the electron temperature
and the divergence of radiative flux within the whole shock wave model.
On the lower panel of Fig.~\ref{itrh} is shown the iterative change of
the total radiative flux $\Fr$ emerging ahead the shock wave.
Thus, if the initial approximation is good enough, we obtained
in all cell centers of the model the
exponentially decaying oscillations of the solution vector.
The global iterations were stopped when relative changes of
variables ceased to decrease. The final relative changes of the solution
vector depend on the both spatial resolution of the discrete model and
accuracy of integration of differential equations
(\ref{dUdX}) -- (\ref{dnidX}). 
It should be noted also that the convergence of global iterations appreciably
varies for different regions of the shock wave. In particular, the convergence
is best in the radiative precursor and thermalization zone, whereas in the
recombination zone becomes slower.
Thus, the convergence plots shown in
Fig.~\ref{itrh} display mostly variations of the solution vector in the
vicinity of the outer boundary of the postshock region.


\section{Results of calculations}

In this paper we discuss the results of calculations done for the shock waves
with upstream velocities $15~\kms\le U_1\le 60~\kms$ propagating
through the unpertubed hydrogen gas of temperature $T_1=3000\K$ and
density $\rho_1=10^{-10}~\gmcs$ ($\nH =6.02\cdot 10^{13}~\mbox{cm}^{-3}$).
In total we computed 46 models with upstream velocity increment
of $\Delta U_1=1~\kms$.
The outer boundary of the preshock region, where the gas is assumed to be
unperturbed, is set at $X\approx -9.2\cdot 10^5$~cm.
Calculations were done for the two-level hydrogen atom, the first
atomic state being treated in non-LTE.
The radiation transfer equation was solved for the both Lyman and Balmer
continua ($\NC=2$).
Thus, the frequency point nearest to the
Lyman edge frequency $\nu_{01} = 3.288\cdot 10^{15}~\hz$
was set at $\nuL=3.754\cdot 10^{15}~\hz$.
More extensive calculations for the larger number of bound atomic states $L$
and hydrogen continua $\NC$ as well as for various temperatures $T_1$ and
densities $\rho_1$ of the unperturbed gas will be given
in the forthcoming paper.

\begin{figure}
\resizebox{\hsize}{!}{\includegraphics{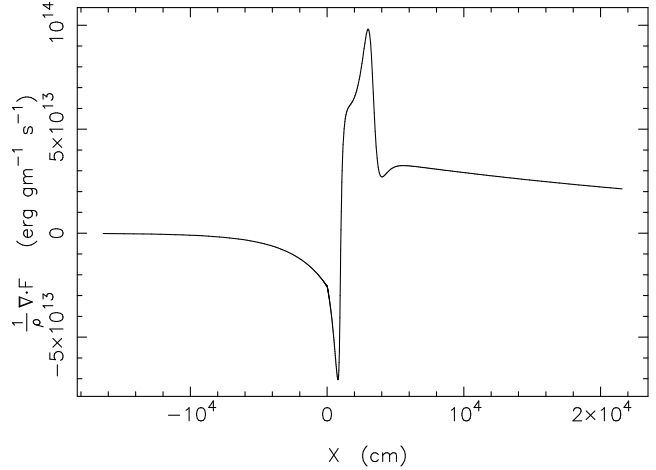}}
\caption{The divergence of radiative flux against
the spatial coordinate in the shock wave model with $U_1=40~\kms$}
\label{u40_drf}
\end{figure}

The most fascinating feature of radiative shock waves is that they
demonstrate the strong interaction between gas material flows and the
radiation field which they produce. This interplay is best seen from the plots
of the divergence of radiative flux as a function of the spatial
coordinate $X$. One of such plots is shown in Fig.~\ref{u40_drf}.
By definition, the divergence of radiative flux is negative, if the fluid
absorbs more energy than emits and, therefore, is heated. And conversely,
when $\df > 0$, the gas radiatively cools since it radiates more energy than
it absorbs.

As is seen in Fig.~\ref{u40_drf}, the divergence of radiative flux is always
negative in the preshock zone, the departure of $\df$ from zero gradually
increasing while the gas approaches the discontinuous jump.
Heating of the precursor
gas material is due to absorption of the Lyman continuum radiation, hence
the region, where the divergence of radiative flux $\df$ perceptibly
deviates from zero, extends over a few units of the optical depth
at frequency $\nuL$.

The properties of the radiation field does not change appreciably across
the discontinuous jump. The spatial resolution of our shock wave models
near the discontinuous jump
is limited by the space interval $\XJ - \XJb = 0.5$~cm. The change of
$\Fr$ and $\drf$ over this interval increases with increasing upstream
velocity $U_1$ but does not exceed 0.3\% for $U_1 = 60~\kms$.
Thus, just behind the discontinuous jump the divergence of radiative flux
goes on to gradually decrease and reaches the minimum
in the vicinity of the maximum of the electron temperature $\Te$.
The rapid growth of the divergence of radiative flux behind its minimum
implies that the gas material flows into the radiatively cooling zone.
For the shock wave model with $U_1=40~\kms$
the distance between maximum and minimum of $\df$ is
$\approx 2.2\cdot 10^3$~cm and corresponds to the optical depth
between these layers of $\tau(\nuL)\approx 3.5$.
It should be noted that both the minimum and maximum of
the divergence of radiative flux as well as other properties
of the shock wave very strongly (for some variables nearly exponentially)
depend on the upstream velocity $U_1$.

\subsection{The radiative precursor}

At the outer boundary of the preshock region the first level of
the hydrogen atom is approximately in LTE since the Lyman continuum
radiation emerging from the postshock region is negligible.
Both the Saha-Boltzmann relation and equations of statistical equilibrium
give nearly the same (within a few per cent)
the hydrogen ionization degree of $\xH\sim 10^{-8}$.
The main sources of opacity are bound-free transitions in the
Lyman continuum (i.e. at frequencies
$\nu\ge\nu_{01}=3.288\cdot 10^{15}~\hz$)
and at lower frequencies
the Rayleigh scattering by hydrogen atoms in the ground state.
The free-free opacity and the Thomson scattering are negligible.
With approaching to the discontinuous jump the hydrogen ionization
degree increases, so that the both free-free opacity and Thomson
scattering increase but nevertheless even for $U_1=60~\kms$
they remain negligible within the entire preshock region.

\begin{figure}
\resizebox{\hsize}{!}{\includegraphics{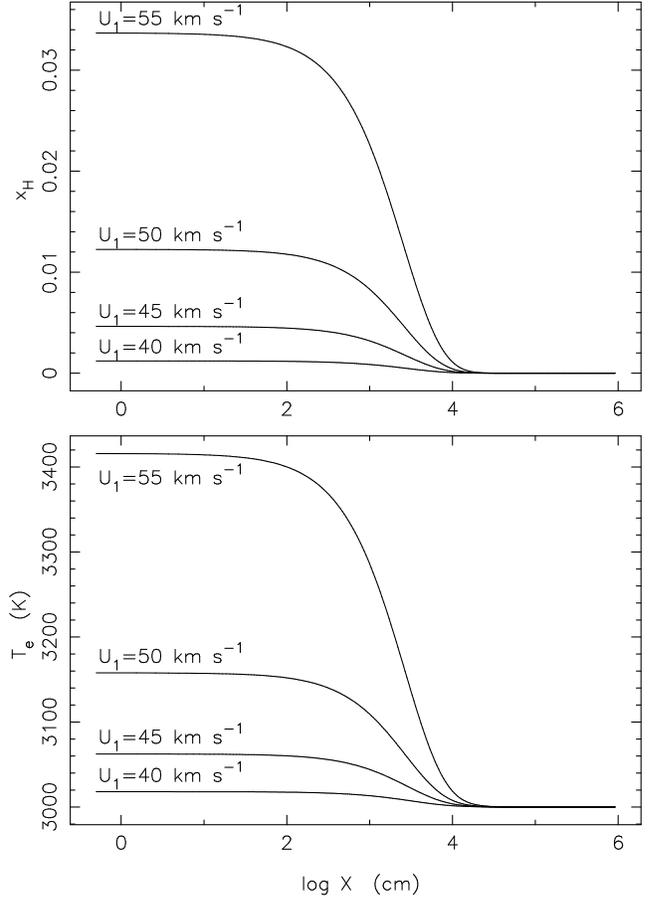}}
\caption{The hydrogen ionization degree $\xH$ (upper panel) and the
electron temperature $\Te$ (lower panel) in the preshock region of
the shock wave models with upstream velocities $U_1 = 40$, 45, 50 and
$55~\kms$}
\label{prec}
\end{figure}

The radiative precursor is revealed as the part of the preshock region
where the hydrogen gas is heated and is ionized by the Lyman continuum
radiation emerging from the postshock region.
The temperature of heavy particles (neutral hydrogen atoms and hydrogen 
ions) $\Ta$ remains constant ($\Ta=T_1=3000\K$) in the radiative precursor.
In Fig.~\ref{prec} are shown the plots of the electron
temperature $\Te$ and the hydrogen ionization degree $\xH$
as a function of the distance from the discontinuous jump
for the shock wave models with upstream velocities
$U_1 = 40$, 45, 50 and $55~\kms$.
As is seen, the perceptible heating and ionization
occur at distances from the discontinuous jump smaller than
$X_{\rm prec}\approx 3\cdot 10^4~\cm$.
The distance $X_{\rm prec}$ corresponds to the optical depth
of $\tau(\nuL)\sim 3$. Thus, for $U_1\le 60~\kms$ the geometrical
thickness of the radiative precursor approximately does not depend
on the upstream velocity.

\begin{figure}
\resizebox{\hsize}{!}{\includegraphics{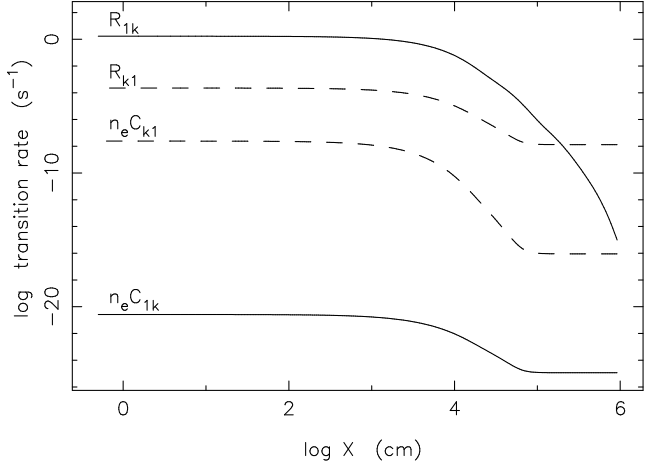}}
\caption{The rates of ionizations (solid lines) and recombinations
(dashed lines) against the distance from the discontinuous jump in
the pre--shock region of the shock wave model with $U_1=40~\kms$}
\label{prec_tr}
\end{figure}

Within the radiative precursor the hydrogen gas and radiation field are in
strong departure from LTE. In particular, ionization of hydrogen atoms is
mainly due to radiative transitions, whereas collisional recombinations
appreciably exceed collisional ionizations.
Fig.~\ref{prec_tr} shows the rates of radiative transitions
$R_{\rm 1k}$ and $R_{\rm k1}=(n_1^*/\ne^*)R_{\rm 1k}^\dagger$
as well as the collisional rates $\ne C_{\rm 1k}$ and
$\ne C_{\rm k1}=$ $\ne (n_1^*/\ne^*)C_{\rm 1k}$ as a function
of $X$ for the shock wave model with $U_1=40~\kms$.
The precursor transition rates are very sensitive to the upstream velocity
$U_1$. For example, for upstream velocities
$20~\kms\le U_1\le 60~\kms$ the rates of ionizations
from the ground state increase
in the ranges $-7.1\le\log R_{\rm 1k}\le 2.6$ and
$-25.8\le\log\left(\ne C_{\rm 1k}\right)\le -10.6$, respectively.
However, notwithstanding such a strong dependence on $U_1$,
for all shock wave models
the spatial dependencies of transition rates were found to be qualitatively
similar to those  displayed in Fig.~\ref{prec_tr}.

The growth of the hydrogen ionization degree with approaching to the
discontinuous jump is accompanied by the increase of the gas pressure
gradient. The corresponding decrease of the gas material velocity becomes,
however, perceptible only for $U > 50~\kms$.
For example, at the inner boundary of the preshock region
the relative decrease of $U$ is $\approx 0.15\%$ for $U_1=55~\kms$
and is $\approx 0.5\%$ for $U_1=60~\kms$.

Of great interest are the physical conditions at the inner boundary of
the preshock region. For the models considered in the present study
the spatial coordinate of this boundary is $\XJb=-0.5$~cm.
For upstream velocities $U_1\la 28~\kms$ both the hydrogen ionization
degree and the electron temperature nearly do not change within the
preshock region. These quantities show the perceptible dependence on
upstream velocity only for $U_1 > 28~\kms$.
In Table~\ref{tibp} we give
the hydrogen ionization degree $\xH$, the electron temperature $\Te$ as well
as the transition rates at the inner boundary of the preshock region.
For the sake of convenience the upstream velocity $U_1$ is given in $\kms$.

\begin{table}
\caption[]{The properties of the preshock inner boundary}
\label{tibp}
\begin{flushleft}
\begin{tabular}{lrrrrrr}
\noalign{\smallskip}
\hline
\noalign{\smallskip}
 & & &
\multicolumn{4}{c}{$\log\left(\mbox{transition rate}~\mbox{s}^{-1}\right)$} \\
$U_1$ & $\log\xH$ & $\Te$
      & $R_{\rm 1k}$ & $R_{\rm k1}$
      & $\ne C_{\rm 1k}$ & $\ne C_{\rm k1}$ \\
\noalign{\smallskip}
\hline
\noalign{\smallskip}
 15 & $-8.04$ & 3000 & $-7.36$ & $-8.79$ & $-25.85$ & $-17.86$ \\
 20 & $-8.04$ & 3000 & $-7.11$ & $-8.79$ & $-25.85$ & $-17.86$ \\
 25 & $-8.04$ & 3000 & $-7.01$ & $-8.79$ & $-25.85$ & $-17.86$ \\
 30 & $-7.21$ & 3000 & $-4.86$ & $-7.99$ & $-25.05$ & $-16.26$ \\
 35 & $-4.19$ & 3002 & $-1.20$ & $-4.93$ & $-21.98$ & $-10.16$ \\
 40 & $-2.92$ & 3018 & $ 0.23$ & $-3.65$ & $-20.58$ & $ -7.61$ \\
 45 & $-2.33$ & 3062 & $ 0.90$ & $-3.03$ & $-19.67$ & $ -6.45$ \\
 50 & $-1.91$ & 3158 & $ 1.40$ & $-2.53$ & $-18.56$ & $ -5.62$ \\
 55 & $-1.47$ & 3416 & $ 1.90$ & $-1.91$ & $-16.46$ & $ -4.77$ \\
 60 & $-0.84$ & 4590 & $ 2.63$ & $-0.75$ & $-10.61$ & $ -3.62$ \\
\noalign{\smallskip}
\hline
\end{tabular}
\end{flushleft}
\end{table}

The electron temperature just ahead the discontinuous jump increases
with increasing upstream velocity nearly exponentially.
For the shock wave models of the present study this dependence can
be approximately expressed due to the following fitting formula
\begin{eqnarray}
\label{prec_te}
\log\Te &=& 3.303 + 2.115\cdot 10^{-2}U_1 - 7.590\cdot 10^{-4}U_1^2 +
\nonumber\\
&+& 8.326\cdot 10^{-6}U_1^3
\,,
\end{eqnarray}
where the upstream velocity $U_1$ is expressed, for the sake of convenience,
in $\kms$.

As is seen in Table~\ref{tibp}, the dominating process just ahead the
discontinuous jump is photoionization and, the\-refore, the shortest relaxation
time in the radiative precursor is that of photoionizations from the
ground state $\trad = 1/R_{1{\rm k}}$. The photoionization relaxation time
gradually decreases with increasing upstream velocity from
$\trad\approx 2\cdot 10^7$~s for $U_1=15~\kms$ to
$\trad\approx 2\cdot 10^{-3}$~s for $U_1=60~\kms$.
Comparing these relaxation times with the time needed for gas to flow through
the precursor $t_{\rm hyd} = X_{\rm prec}/U_1$ we find that
for models with upstream velocities $U_1 <50~\kms$ the ratio
of the photoionization relaxation time to the hydrodynamic time is
$\trad/t_{\rm hyd} \gg 1$ and only for $U_1=60~\kms$
the photoionization relaxation time becomes nearly comparable with
hydrodynamic time: $\trad/t_{\rm hyd}\approx 0.4$.
Because for establishment of the statistical equilibrium
this ratio should be $\trad/t_{\rm hyd}\ll 1$,
the ground state populations of the hydrogen atom are in
strong departure from the statistical equilibrium
and the hydrogen ionization degree
cannot be described in assumption of statistical equilibrium.
For example, at the inner boundary of the preshock region of the
shock wave model with $U_1=50~\kms$ the hydrogen ionization degree
is $\xH=1.22\cdot 10^{-2}$, whereas solution of the equations
of statistical equilibrium gives $\xH=0.99$.

\subsection{The thermalization and recombination zones}

Behind the discontinuous jump the translational kinetic energy of heavy
particles is redistributed among various degrees of freedom characterized
by different relaxation times. The fastest relaxation process is the
translational kinetic energy exchange in elastic collisions of electrons
with neutral atoms and ions. Another relaxation process
is excitation of bound atomic states and ionization of hydrogen atoms.
Both excitation and ionization need, however, much more
collisions than translational kinetic energy exchange
(Stupochenko et al.~1967), so that just behind the
discontinuous jump the electron temperature gradually increases, whereas
the hydrogen ionization degree remains nearly constant.
Note that although the bound--bound transitions were not considered
in our model, the excitation of atomic states is taken into
account as a result of bound--free transitions.

\begin{figure}
\resizebox{\hsize}{!}{\includegraphics{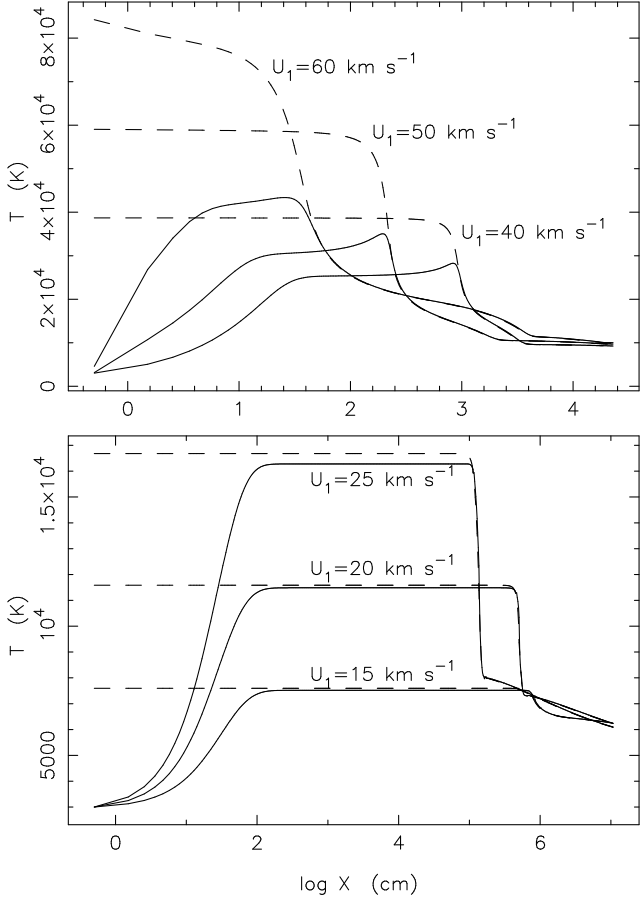}}
\caption{The electron temperature $\Te$ (solid lines) and temperature
of heavy particles $\Ta$ (dashed lines) as a function of distance from
the discontinuous jump in the postshock region}
\label{wake_t}
\end{figure}

In Fig.~\ref{wake_t} are shown the electron temperature $\Te$ and the
temperature of heavy particles $\Ta$ as a function of distance from the
discontinuous jump for shock wave models with upstream velocities from
$15~\kms$ to $60~\kms$. As is seen, the characteristic time of the electron
temperature growth rapidly decreases with increasing upstream
velocity. Furthermore, for $U_1 < 30~\kms$ there is a temperature plateau
within of which both temperatures approximately do not change until they
begin to decrease.
The existence of the temperature plateau is due to the fact that
for upstream velocities $U_1 < 30~\kms$ the electrons acquire the
energy from heavy particles due to elastic collisions with neutral
hydrogen atoms.
The rate of energy gain in such collisions is very small and gradually
decreases when the electron temperature $\Te$ approaches the
temperature of heavy particles $\Ta$.
The temperature plateau appears when the electron energy gain in elastic
collisions with neutral hydrogen atoms becomes almost negligible.
At upstream velocities $U_1 > 30~\kms$ the electron
temperature plateau is ended by the slight bump, the bump being wider and
higher with increasing $U_1$.

In the second and third columns of Table~\ref{tpsr} are given
both the maximum value of the electron
temperature in the postshock region as well as the distance of this
point from the discontiuous jump expressed in cm.
In the first column of Table~\ref{tpsr} we give the upstream velocity
expressed, for the sake of convenience, in $\kms$.
The electron temperature peak
was not found for the shock wave models with $U_1\le 20~\kms$ and
in these cases we give only the electron temperature of the plateau.
The electron temperature maximum can be approximately
considered as a point where temperatures of heavy particles and free
electrons equilize. Thus, the width of the relaxation zone
where both temperatures equalize decreases by a factor of
$\approx 3600$ for the upstream velocity increasing from
$25~\kms$ to $60~\kms$.
It is of interest to note that for upstream velocities
$U_1 < 50~\kms$ the temperature of heavy particles
$\Ta$ remains nearly constant until the electron temperature begins to
increase just before its drop.
This is due to the fact that at small upstream velocities the
fractional abundance of free electrons is so small that they
cannot perceptibly affect the gas of heavy particles.

\begin{table}
\caption[]{Properties of the postshock region}
\label{tpsr}
\begin{flushleft}
\begin{tabular}{lrccccc}
\noalign{\smallskip}
\hline
\noalign{\smallskip}
$U_1$ & $\Te$ & $\log X$
      & $\xH$ & $\log X$
      & $\!\log\drf\!$ & $\!\log{\FLy\over\Fr}\!$ \\
\noalign{\smallskip}
\hline
\noalign{\smallskip}
 15 & 7521 &      & 0.014 & 6.29 & 11.097 & $-5.22$ \\
 20 &11486 &      & 0.071 & 5.72 & 12.671 & $-3.43$ \\
 25 &16281 & 4.95 & 0.134 & 5.16 & 13.201 & $-2.83$ \\
 30 &20986 & 4.20 & 0.206 & 4.40 & 13.473 & $-2.80$ \\
 35 &24712 & 3.42 & 0.294 & 3.89 & 13.749 & $-2.19$ \\
 40 &28274 & 2.92 & 0.406 & 3.61 & 13.992 & $-1.81$ \\
 45 &31681 & 2.58 & 0.528 & 3.43 & 14.198 & $-1.66$ \\
 50 &35050 & 2.29 & 0.662 & 3.32 & 14.381 & $-1.49$ \\
 55 &38575 & 1.96 & 0.805 & 3.22 & 14.569 & $-1.23$ \\
 60 &43371 & 1.40 & 0.964 & 2.89 & 14.848 & $-0.74$ \\
\noalign{\smallskip}
\hline
\end{tabular}
\end{flushleft}
\end{table}

\begin{figure}
\resizebox{\hsize}{!}{\includegraphics{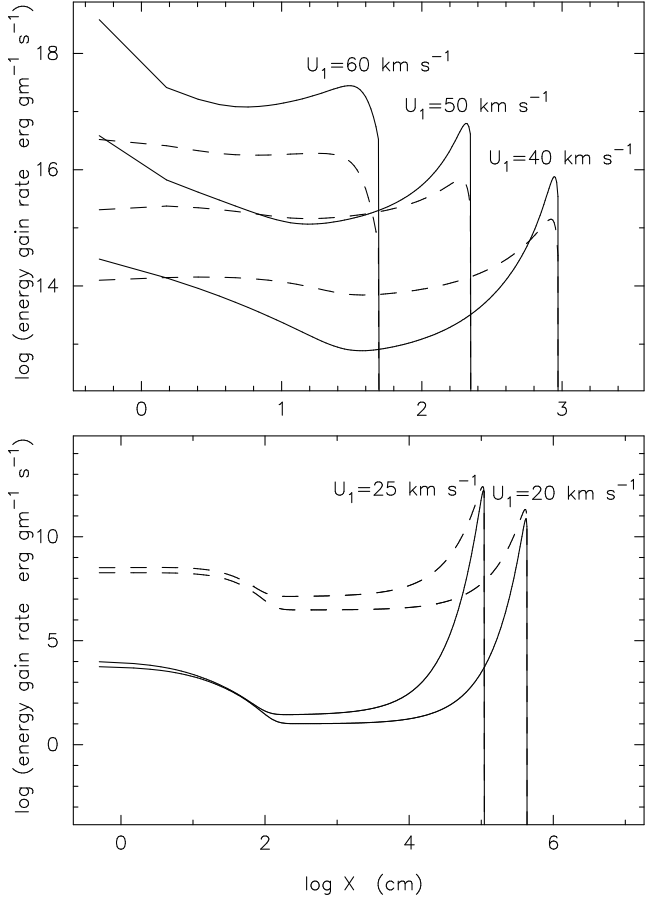}}
\caption{The rates of energy gain by electrons in elastic collisions with
hydrogen ions (solid lines) and neutral hydrogen atoms (dashed lines)}
\label{wake_q}
\end{figure}

The translational energy exchange between heavy particles and electrons
is due to elastic collisions of electrons with both neutral hydrogen atoms
and hydrogen ions. The cross section of elastic collisions
with hydrogen ions is much larger than that of electrons with neutral atoms
and, therefore, the translational energy gain by electrons from heavy
particles is strongly dependent on the hydrogen ionization degree.
The rates of energy gain by electrons in both these processes are shown
in Fig.~\ref{wake_q}. The abrupt decrease of $\Qei$ and $\Qea$ occurs
when both temperatures equalize.

For shock wave models considered in the present study the rate of energy gain
by electrons in elastic collisions with ions exceeds that
in elastic collisions with neutral atroms only when the hydrogen
ionization degree is $\xH \ga 10^{-2}$.
As is seen in Fig.~\ref{wake_q}, for upstream velocities $U_1\le 25~\kms$,
the electron temperature equalizes with temperature of heavy particles
only due to elastic collisions of electrons with neutral atoms.
Comparing with lower panel of Fig.~\ref{wake_t} one sees that
when the electron temperature reaches the plateau, the rate of energy
gain $\Qea$ decreases by nearly two orders of magnitude.

At upstream velocities $30~\kms < U_1 < 50~\kms$, the energy gain
in elastic collisions with ions begins to dominate only just before
the temperature drop. It is this increase of $\Qei$ that is responsible
for the electron temperature peak near the end of the electron temperature
plateau. The translational energy gain by
electrons in elastic collisions with hydrogen ions becomes completely
dominating for $U_1 > 50~\kms$.
The gradual decrease of $\Qei$ just behind the discontinuous jump in the
shock wave models with $U_1\ge 40~\kms$ (see upper panel of Fig.~\ref{wake_q})
is due to the temperature dependence of the equipartition time given in
Eq.~(\ref{teq}).

\begin{figure}
\resizebox{\hsize}{!}{\includegraphics{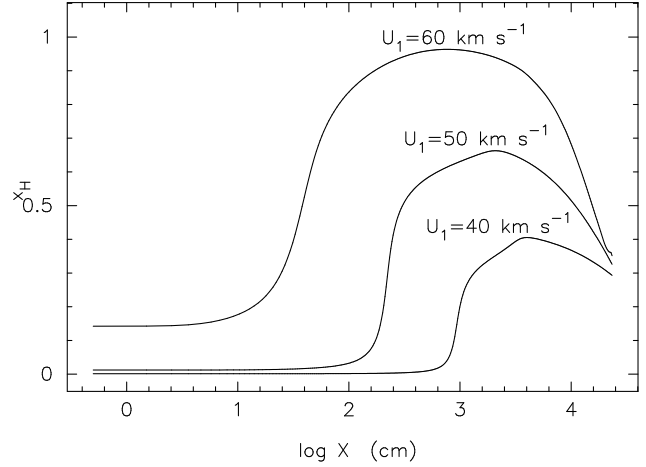}}
\caption{The hydrogen ionization degree as a function of the distance
from the discontinuous jump in the shock wave models with upstream velocities
$U_1 = 40$, 50 and $60~\kms$}
\label{wake_x}
\end{figure}

In Fig.~\ref{wake_x} are shown the plots of the hydrogen ionization degree
in the postshock region of the shock wave models with upstream velocities
$U_1 = 40$, 50 and $60~\kms$.
Comparing with Fig.~\ref{wake_t} one sees that
the maximum of the hydrogen ionization occurs at much larger distances
from the discontinuous jump than the maximum of the electron temperature.
Very approximately the distances of both these maxima can be expressed
as power functions of the upstream velocity:
\begin{eqnarray}
\label{maxTe}
X(\max\Te) &=& 1.21\cdot 10^9 U_1^{-6.96}
\,,\\
\label{maxxH}
X(\max\xH) &=& 6.07\cdot 10^7 U_1^{-5.60}
\,.
\end{eqnarray}
In these approximate expressions, for the sake of convenience,
the distances $X(\max\Te)$ and $X(\max\xH)$
are expressed in km whereas the upstream velocity is expressed in $\kms$.
The maximum values of $\xH$ as well as the corresponding distances from the
discontinuous jump expressed in cm are given in the fourth and fifth colums of
Table~\ref{tpsr}, respectively.
According to the schematical division of the shock wave noted above
the maximum of $\xH$ can be considered as the boundary between
the thermalization and recombination zones.

\begin{figure}
\resizebox{\hsize}{!}{\includegraphics{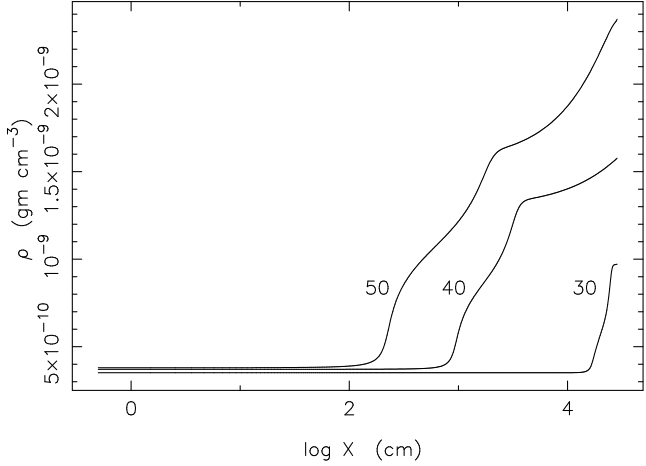}}
\caption{The postshock gas density $\rho$ against the distance
from the discontinuous jump. The numbers at the curves indicate
the upstream velocity $U_1$ in $\kms$}
\label{wake_rho}
\end{figure}

The large distance between maxima of the electron temperature
and the hydrogen ionization degree implies that the degree
of freedom associated with ionization of hydrogen atoms is frozen
in comparison with excitation of translational motions.
As a result, the gas flows through the maximum of $\Te$ at
smaller heat capacity $C_{\rm p}$ than it would be in equilibrium.
Thus, when the perceptible fraction of hydrogen atoms is ionized,
the gas density begins to increase. At larger distances the gas density
goes on to increase due to the radiative cooling
(see Fig.~\ref{wake_rho}).

\begin{figure}
\resizebox{\hsize}{!}{\includegraphics{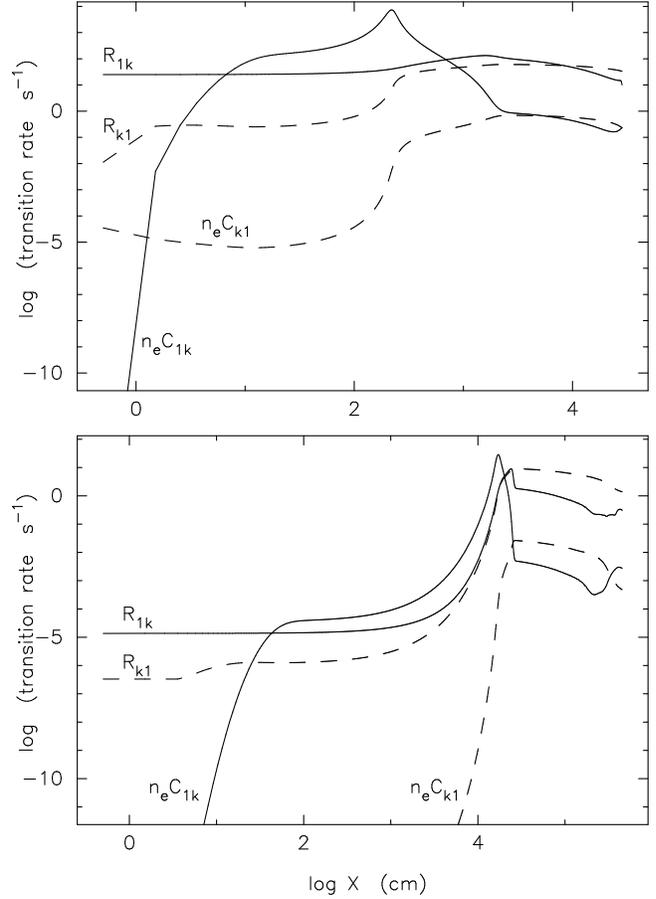}}
\caption{The rates of ionizations (solid lines) and recombinations
(dashed lines) against the distance from the discontinuous jump $X$ in the
postshock region. Upper panel: $U_1=50~\kms$; lower panel: $U_1=30~\kms$}
\label{wake_tr}
\end{figure}

The total ionization rate of hydrogen atoms is a sum of photoionizations
and collisional ionizations, so that it depends on the number density
of free electrons. Because no ionization occurs across the discontinuous
jump, the total ionization rate behind the discontinuous jump strongly
depends on the ionization in the radiative precursor.
To demonstrate this dependence, in Fig.~\ref{wake_tr} are shown the
rates of ionizations and recombinations, both collisional and radiative,
in the postshock regions of the shock waves with upstream velocities
30 and $50~\kms$.

For the shock wave model with $U_1=30~\kms$ the hydrogen ionization
degree at the discontinuous jump is $\xH\approx 6\cdot 10^{-8}$,
therefore just behind the discontinuous jump the role of collisional
ionizations is negligible and the number density of free electrons
increases initially only due to photoionizations.
For $U_1=50~\kms$ the hydrogen ionization degree at the
discontinuous jump is $\xH\approx 10^{-2}$.
Free electrons created in the precursor play a role of the seeds
producing yet more electrons and leading to the electron avalanche.

Compared to the preshock region where the contribution of opacity
sources nearly does not depend on the upstream velocity,
the gradual growth of the hydrogen ionization degree behind the discontinuous
jump leads to the appreciable dependence of postshock opacities
on the upstream velocity $U_1$.
For $U_1 < 30~\kms$ the postshock opacity in the Balmer continuum is mainly
due to Rayleigh scattering by neutral hydrogen atoms. Only
near the Balmer edge, the opacity due to bound-free transitions from 
the second level becomes most important.
With increasing upstream velocity the role of the
bound-free transitions from the second level increases
and for $U_1 > 50~\kms$ this opacity mechanism becomes dominating
within the whole range of the Balmer continuum.
The role of the Thomson scattering is somewhat perceptible only
near the Balmer edge, whereas the free-free opacity can be neglected.

\subsection{The radiation field}

Although we did not include into
the system of ordinary differential equations
(\ref{dUdX}) -- (\ref{dnidX}) the terms with radiation
energy density $\Er$ and radiation pressure $\Pr$, these quantities were
evaluated as
\begin{equation}
\label{Er}
\Er = \frac{4\pi}{c}\int\limits_0^\infty \int\limits_0^1\u d\mu d\nu
\end{equation}
\begin{equation}
\label{Pr}
\Pr = \frac{4\pi}{c}\int\limits_0^\infty \int\limits_0^1 \u\mu^2 d\mu d\nu
\end{equation}
together with total radiative flux $\Fr$ and divergence of
the radiative flux $\df$ each time when the radiation transfer equation
was solved. For the fixed spatial coordinate $X$ the ratio of the
radiation energy density to the total translational kinetic energy
gradually increases with increasing upstream velocity and is
in the range $5\cdot 10^{-4}\la\Er/(\Ea+\Ee)\la 0.02$ for
$20~\kms\le U_1\le 60~\kms$. The ratio $\Er/(\Ea+\Ee)$
is highest at the inner boundary of the preshock region
because just behind the discontinuous jump the total translational kinetic
energy increases by more than an order of magnitude, whereas the
total change of $\Er$ within the shock wave does not exceed 30\%.
Because the most of the energy flux is contained in the radiative
flux $\Fr$, we find that $U\Er/\Fr\la 2.3\cdot 10^{-4}$
for $U_1=60~\kms$.
Thus, our assumption that the radiation energy and radiation pressure can be
neglected is enough good.

\begin{figure}
\resizebox{\hsize}{!}{\includegraphics{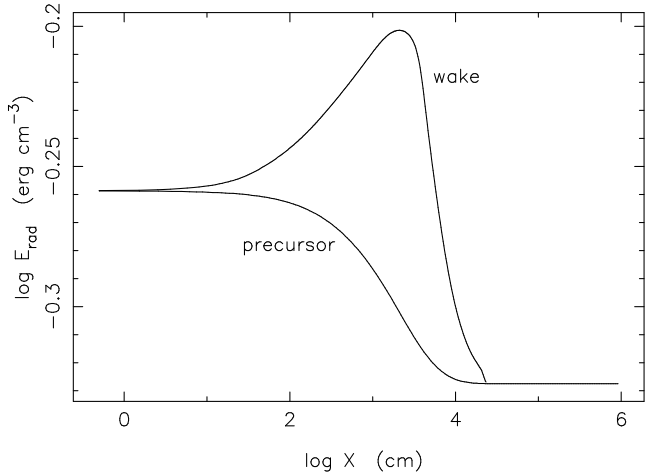}}
\caption{The total radiation energy density as a function of distance from
the discontinuous jump in the shock wave model with $U_1=60~\kms$}
\label{slab_er}
\end{figure}

In Fig.~\ref{slab_er} are shown the dependencies of $\Er$ on the distance
from the discontinuous jump for the both pre\-shock and postshock regions
of the model with $U_1=60~\kms$. The increase of $\Er$ in the preshock
region with approaching to the discontinuous jump is due to the radiative
heating by the layers absorbing the Lyman continuum radiation.
Behind the discontinuous jump the radiation energy density reaches the
maximum in the hydrogen recombination zone.

The most of the shock wave radiation, by definition,
is produced in the layers where the
divergence of radiative flux $\df$ reaches the maximum.
These layers locate nearly at the same distance from the discontinuous
jump as the maximum of the hydrogen ionization degree. The maximum of
$\drf$ exponentially increases with increasing upstream velocity and
for $20~\kms\le U_1\le 60~\kms$ is given by

\begin{equation}
\left(\drf\right)_{\max} = 1.25\cdot 10^{12}\exp\left(0.105U_1\right)
\ \mbox{erg}~\mbox{gm}^{-1}\mbox{s}^{-1},
\end{equation}
where the upstream velocity is expressed in $\kms$.
Maximum values of $\drf$ are given in the sixth column of Table~\ref{tpsr}.

\begin{figure}
\resizebox{\hsize}{!}{\includegraphics{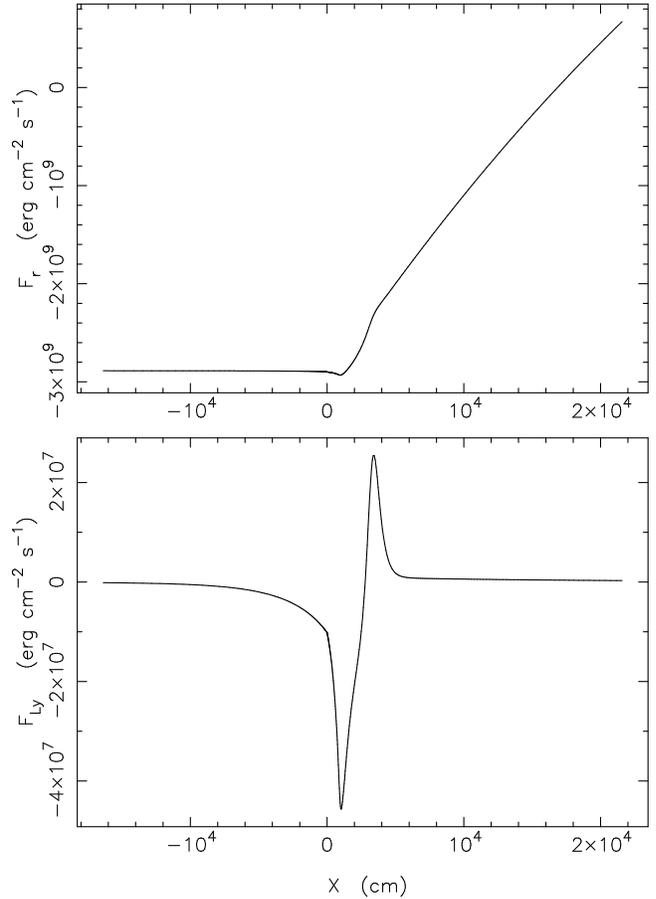}}
\caption{The radiative flux within the Lyman continuum $\FLy$ (lower panel)
and the total radiative flux $\Fr$ (upper panel) against the spatial
coordinate $X$ for the shock wave model with $U_1=40~\kms$}
\label{u40_flux}
\end{figure}

As was noted above the shock wave models have a small optical depth in
the Balmer continuum
($\tau_{\max}\sim 10^{-4}$) and at the same time they are opaque
for the Lyman continuum photons. Therefore, the radiation emerging from
the both surfaces of the slab is mostly within the Balmer continuum
whereas the Lyman continuum radiation is transported only within the narrow
zone surrounding the discontinuous jump.
In the lower panel of Fig.~\ref{u40_flux} for the shock wave model with
upstream velocity $U_1=40~\kms$
is shown the Lyman continuum radiative flux $\FLy$
as a function of the spatial coordinate $X$.
This dependence is qualitatively similar to that of
the divergence of radiative flux shown in Fig.~\ref{u40_drf}.
The negative flux corresponds to the radiation propagating in the
direction of the preshock region.
The optical depth between minimum and maximum of $\FLy$
is $\Delta\tau(\nuL)\approx 4.4$ for $U_1=40~\kms$
and slowly decreases from $\Delta\tau(\nuL)\approx 24$ for $U_1=30~\kms$
to $\Delta\tau(\nuL)\approx 1$ for $U_1=60~\kms$.
This decrease is mostly due to the growth of the hydrogen ionization degree
in the postshock region.
The ratios of the maximum radiation flux transported within the Lyman
continuum to the total radiation flux are given in the last column of
Table~\ref{tpsr}.

In the upper panel of Fig.~\ref{u40_flux} for the same model
is shown the total radiative flux $\Fr$.
This plot displays only the very vicinity of the discontinuous jump.
In the pre\-shock region, beyond the radiative precursor,
the total radiative flux $\Fr$ remains constant
since the divergence of radiative flux is $\df=0$.
In the postshock region the total radiative flux gradually increases
and asymtotically tends to the same value as in the preshock but with
the opposite sign.
For all shock wave models we obtained that the total radiative fluxes
emerging from both surfaces of the slab are $-F_{\rm r1} = F_{{\rm r}N}$.


\section{Conclusion}

The primary goal of the work reported in this paper was to obtain
the self-consistent solution of the equations of fluid dynamics,
rate equations and radiation transfer equation for the structure
of the steady shock wave.
The procedure of global iterations described in the present paper
in general resembles the compute of stellar atmosphere models.
Indeed, like in stellar atmosphere calculations the shock wave model
takes into account the coupling between the gas material and
radiation field.
The self-consistent model is obtained with iteration procedure
comprising the solution of the radiation transfer equation
and integration of the mass, momentum and energy conservation equations
written in the form of the ordinary differential equations.
Each cycle of iterations gives, in general, improved characteristics
of the gas and radiation field.

At the same time, the problem of the shock wave structure compared to
that of stellar atmosphere models contains a number of serious
complications. First, atomic level populations are not only in strong
departures from LTE but are also in significant departures
from statistical equilibrium.
Second, unlike the stellar atmospheres, where the divergence of
radiative flux is $\df=0$ (the condition of radiative equilibrium),
in shock waves the part of the energy of hydrodynamic flow is transformed
into radiation and the radiative equilibrium is established only far away
from the discontinuous jump.
Furthemore, in stellar atmosphere models the total radiative flux is
given as one of the boundary conditions, whereas in the shock wave model
the emerging flux is obtained from the solution of the problem.
The small optical depth increments in hydrogen continua of order
$l\ge 3$ lead to the losses of the machine accuracy when the Feautrier
technique is applied. The problem is so serious, that even the improved
method by Rybicki \& Hummer (1991) sometimes fails.
Third, the rate equations are stiff and need the special treatment
in their solution.
In particular, the convergence
of global iterations depends on the tolerance parameter determining
the maximum error permitted during the integration.

The present paper is confined by consideration of the two-level atomic
model, so that the radiation transfer is treated for the Lyman and Balmer
continua, only. This approximation seems to be insufficient for the shock wave
problem because the occupation numbers of levels $l\ge 2$ obviously deviate
from LTE and the perceptible fraction of radiation is transported at
frequencies lower than the Balmer edge frequency. Nevertheless, notwithstanding
such a restriction, there is a qualitative agreement of our results with those
obtained earlier by other authors. For example, according to
calculations of Gillet \& Lafon (1990) the electron temperature just ahead
the discontinuous jump is $\Te\approx 14000\K$
for the upstream velocity of $U_1=80~\kms$.
Although in the present study the highest upstream velocity was
$U_1=60~\kms$, a very approximate comparison can be done with
fitting formula (\ref{prec_te}) which gives the same electron temperature
for the upstream velocity $U_1\approx 75~\kms$.
Thus, more detailed calculations are needed and
in the forthcoming paper we are going to present the grid of the
shock wave models computed for the larger number of hydrogen atomic le\-vels
and wider range of upstream velocities,
the models of the present study being used as
initial approximation for more correct shock wave models.

More realistic models, however, should take into account not only
bound-free terms but also bound-bound terms in the rate equations
and the radiation transfer problem should be solved for the both
continuum and spectral line radiation. This is the perspective 
for the near future.
It is certainly one of the most basic. Indeed, 
preceding shock studies show that radiative processes, which 
determine the wake cooling, have a strong influence on the 
resulting shock structure. Because in the model of this paper 
we consider a pure H-plasma without H$^{-}$, and
only include the bound-free photo- and collisional processes
of H atoms, we expect that the absence of some predominant coolants 
such as neutral and singly ionized metal atoms might appreciably 
underestimate the radiative cooling rate of the gas. 
The importance of radiative heating and cooling rates in 
shocked circumstellar envelopes have been recently investigated
(Woitke et al. 1996) but only a few transitions of the numerous
metal lines were considered. At present such a basic study seems 
to be beyond our immediate abilities and is out of the scope of
our paper which was to investigate the possibility of obtaining
a self-consistent solution of the structure of radiative shock
waves in dense atmospheric gas.

\begin{acknowledgements}
The authors are indebted to the referee, Dr.\ Peter\ Woitke, for his
meticulous review of this paper and for his valuable comments.
The work of YAF has been done in part under the auspices of the
Minist\`ere de l'Enseignement Su\-p\'erieur et de la Recherche
(grant 1196/MDLM/CB) in 1996 and of the Aix-Marseille~I University
(grant 34 PR 1049) in 1997.
YAF acknowledges also the support from the Russian Foundation
for Basic Research (grant 95-02-06359).
\end{acknowledgements}

\end{document}